\documentstyle[epsf,11pt,twoside,flushrt]{article}                         

\catcode`\"=\active\let"=\"
\def\3{\ss }
\newcommand{\halpha}{H$\alpha$}

\newcommand{\ion}[2]{#1\,{\sc #2}}
\newcommand{\teff}{T_{\mathrm{eff}}}         
\def\fsek{\hbox{$.\!\!{''}$}}
\def\sek{\hbox{$\hbox{\thinspace \fsek}$}}
\newcommand{\litanf}{\begin{list}{}{\leftmargin=1.5cm \rightmargin=0cm 
\itemindent=-1.5cm \parsep=0cm \itemsep=0cm }}
\newcommand{\litend}{\end{list}}
\newcommand{\Msolar}{\mbox{\,$\rm M_{\odot}$}}        
\begin{document}                                                      
\noindent{\bf                                                         
THE UV BRIGHT STAR ZNG\,1 IN M\,5
}                                                                     

\begin{list}{}{\topsep 0in                                            
          \partopsep 2\baselineskip                                   
          \itemsep 0pt                                                
          \parsep \baselineskip                                       
          \leftmargin .53in                                           
          \listparindent 0in                                          
          \labelsep 0in                                               
          \labelwidth 0in}                                            
\item ~                                                               


Ralf Napiwotzki, Ulrich Heber

Astronomisches Institut der Universit\"at Erlangen-N\"urnberg,
Dr. Remeis-Sternwarte, Sternwartstr. 7, D-96049 Bamberg, Germany

\end{list}                                                              


\vspace{2\baselineskip}                                                

\noindent{ABSTRACT:}                                                  
We report the results of \halpha\ imaging and UV spectroscopy with the
Hubble Space Telescope of the hot post-AGB star ZNG\,1 in M\,5 and its
suspected planetary nebula. 

\vspace{\baselineskip}                                                

\pagestyle{myheadings}                                                
\markboth{\hspace*{1.0in}{\rm                                         
R. Napiwotzki \& U. Heber
}\hspace{\fill}}{{\rm                                                 
The UV bright star ZNG\,1 in M\,5
}}                                        

\section{UV bright stars}

Post asymptotic giant branch (PAGB) stars of population~II are
rare objects but are Rosetta stones for testing the theory of stellar 
evolution. Bona fide members of this class are the rare nuclei of halo 
planetary nebulae (PN) and PAGB stars in globular clusters. 
The cluster PAGB stars have to evolve from low initial progenitor masses
($\approx$turn-off mass of the cluster $< 1M_\odot$). Thus these PAGB 
objects and their PNe provide us with important tests of the evolution 
of low mass stars. 

Most cluster PAGB stars were detected because of their brightness in the 
UV region. In their pioneering work Zinn et al.\ (1971; ZNG) selected hot
luminous objects in a number of globular clusters by looking for stars of 
exceptional brightness in the Johnson U band and termed these objects 
UV bright stars.
Zinn et al.\ listed seven UV-bright stars for the globular 
cluster M\,5 (NGC\,5904), ZNG\,1 is located very close to the cluster center. 

A low resolution IUE spectrum of ZNG\,1 obtained by Bohlin et al.\ (1983)
showed strong
resonance lines of N\,V (1240\,\AA) and C\,IV (1550\,\AA) and an emission
of the semiforbidden NIV] 1487\,\AA\ line. Bohlin et al.\ (1983) estimated
an effective temperature of $\approx$35000\,K. A comparison with an IUE
spectrum of the main sequence O star $\mu$\,Col revealed that the N\,V
1240\,\AA\ resonance lines are much stronger in the low metallicity star
ZNG\,1 than in the pop.~I star $\mu$\,Col. Bohlin et al.\ explained this
finding by means of the dredge-up of CNO processed material in ZNG\,1.
The NIV] emission was
interpreted as evidence for a possible PNe around this hot post-AGB 
object. 

\section{Imaging and UV spectroscopy}

Optical imaging of ZNG\,1 with the ESO-NTT in the UBV bands 
demonstrated that ZNG1 is double with a G-type companion being only 
0\sek 5 apart, which makes
further investigation of ZNG\,1 by ground based observations virtually
impossible.
Hence we obtained HST observations with the GHRS spectrograph and the 
WPFC2 camera to improve our understanding of this object.

Since up to now only four PNe in globular cluster are known any 
new object is of importance. Thus we obtained a WFPC2 \halpha\ image of 
ZNG\,1 (F656N filter) to check the PN hypothesis of Bohlin et al.\ (1983). 
The central region of the PC image is displayed in Fig.~1. 
Due to the superior angular separation of HST both components of the binary 
are easily resolved.

\begin{figure}[htbp]
\begin{center}
\epsfxsize=10.0cm
\mbox{
\epsffile[70 300 520 750]{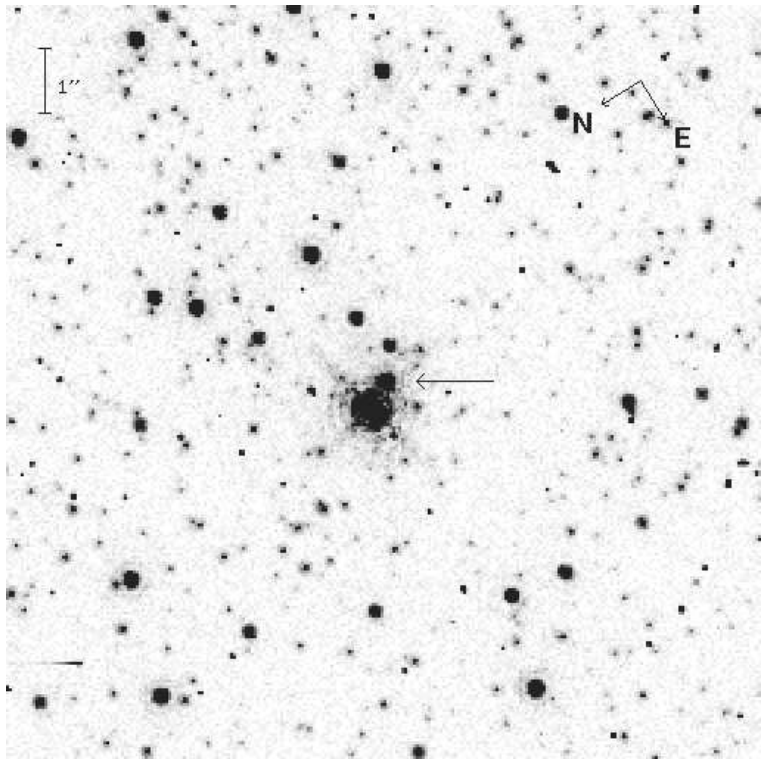}
}
\caption{Enlarged area of the PC F656N image of M\,5, showing ZNG\,1 and 
its companion. The UV bright star is marked with an arrow.}
\end{center}
\end{figure}

Ultraviolet spectra of ZNG\,1 were obtained with the Goddard high 
resolution spectrograph onboard of the Hubble Space Telescope using the 
grating G140L. Two spectra were taken through the large science aperture which 
covered the wavelength range 1150\,\AA\ to 1430\,\AA\ and 1470\,\AA\ and 
1750\,\AA, respectively. The achieved spectral resolution is 0.7\,\AA.
The spectra are displayed together with line identifications in Fig.~2
for both wavelengths ranges. Most remarkably there is no trace of the 
semiforbidden NIV] 1487\,\AA\ line. We conjecture that its identification 
in the IUE spectrum was caused by an artefact.

\begin{figure}
\begin{center}
\epsfxsize=15.0cm
\mbox{
\epsffile[50 30 740 460]{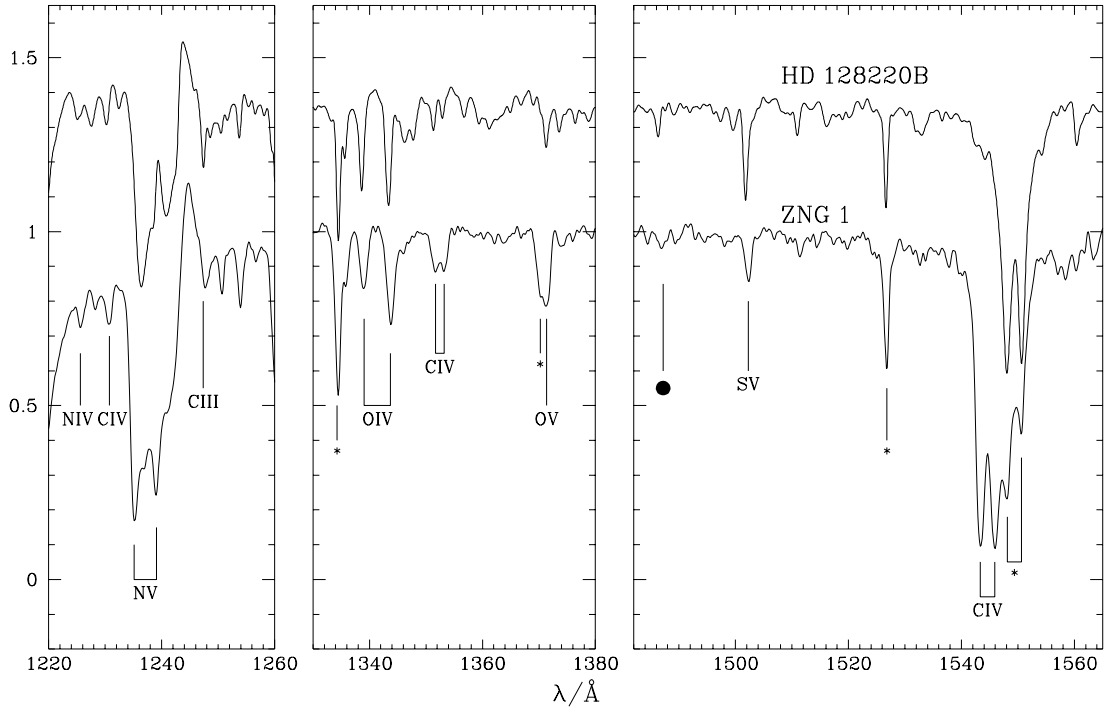}
}
\caption{GHRS spectrum of ZNG\,1, compared to the IUE spectrum of 
the sdO star HD\,128220\,B (top). Left hand panel: region around the \ion{N}{v}
doublet, middle panel: \ion{O}{iv} and \ion{O}{v} lines, right hand panel:
region around the \ion{C}{iv} resonance doublet.
Important photospheric lines and wind features
are identified, interstellar lines are marked with an asterisk. The position
of the claimed \ion{N}{iv]}\,1487\,\AA\ emission is marked by a filled
circle.}
\end{center}
\end{figure}

The UV spectra of ZNG1 are compared in Fig.~2 to the IUE spectra of the
 well known sdO 
star HD\,128220\,B (superposition of 30 IUE high resolution spectra, 
degraded to 0.7\AA\ resolution). There are striking 
similarities: The \ion{N}{V} resonance doublet is a P-Cygni profile in both 
stars indicating that it is formed in a stellar wind. The \ion{C}{IV} line
is shifted bluewards by 900 km/s in ZNG\,1 but unshifted in HD\,128220\,B, 
indicating that it is formed in a stellar wind in case of ZNG\,1 but is of
atmospheric origin in HD\,128220\,B (note that the \ion{C}{IV} resonance lines 
of HD\,128220\,B
displays time variable wind features, i.e. narrow components. Only spectra 
without \ion{C}{IV} wind features were coadded, see Rauch, 1993).
Photospheric lines of \ion{C}{III},\ion{C}{IV},\ion{N}{III},
\ion{N}{IV}, \ion{O}{IV}, \ion{O}{V} and \ion{S}{V} can be identified in 
both spectra with line strengths remarkably similar (see Fig. 2).

The atmospheric UV line spectrum of HD\,128220\,B has been analysed 
extensively by Rauch (1993), who derived the atmospheric parameters
$\teff$=40600K$\pm$0.4K, log\,g = 4.5$\pm$0.1. The atmosphere is enriched in 
helium (He/H=0.3) and nitrogen (1.0 dex) with respect to solar abundance, 
whereas oxygen is depleted (by 0.64 dex) and carbon is almost solar. The 
\ion{N}{V} P-Cygni profile was analysed by Hamann et al. (1981) and a 
mass loss rate in the range 10$^{-10.9}$ -- 10$^{-8.5}$ \Msolar/yr was derived.

\section{Conclusions}

ZNG\,1 and its cool companion are easily resolved in our PC image. Their 
separation is 0\sek 5. Adopting a distance for M\,5 of 7600\,pc 
(Harris \& Racine 1979) we 
derive a projected physical distance of 3700\,A.U. between ZNG\,1 and the bright
companion.
Since it is very likely that such a wide pair would be disrupted in the dense
environment of a globular cluster, we conclude that these stars don't form
a binary but are a chance alignment.
No nebula was detected on our \halpha\ image of ZNG\,1 (see Fig.~1). Since the
stellar image of ZNG\,1 looks like any other star we conclude that we can
rule out any PN more extended than, say, 0\sek 2 ($7\cdot 10^{-3}$\,pc). 
Neither could the presence of the semiforbidden NIV] 1487\,\AA\ line be 
confirmed.

The richness and strength of the C, N and O line spectrum of ZNG 1 is surprising
since the star resides in metal poor environment (Fe/H=-1.4, Djorgovski
1993). From its spectral similarities with the well studied sdO star 
HD\,128220\, it is tempting to 
conjecture that C, N and O have been enriched by nuclear burning and 
subsequently mixed to the stellar surface. A similar case is reported for 
K\,648, the central star of a planetary nebula in the globular cluster M\,15, 
which is enriched in carbon by more than a factor of 100 with respect to 
the cluster metallicity (Heber et al., 1993). The results of a quantitative 
spectral analysis must be awaited to prove or disproof our conjecture. 

{\it Acknowledgments.}
The help of M.~Rosa in preparing our HST observations is gratefully 
acknowledged. R.N.\ is supported by the DARA under grant 50\,OR\,9309.
Attendance of U.H. at the FBS meeting was made possible by a DFG travel grant.

\bigskip                                                              
\pagebreak[1]                                                         
\noindent REFERENCES:                                                 
\nopagebreak[4]                                                       
\begin{list}{}{\topsep -\baselineskip                                 
		\partopsep 0in                                        
		\itemsep 0in                                          
		\parsep 0in                                           
		\leftmargin .35in                                     
		\listparindent -.35in                                 
		\labelsep 0in                                         
		\labelwidth 0in}                                      
\item ~

Bohlin R.C., Cornett R.H., Hill, J.K., Smith A.M., Stecher T.P.,
Sweigart A.V. 1983, ApJ 267, L89


Djorgovski S.G. 1993, PASPC 50, 373

Harris W.E., Racine R. 1979, ARAA 17, 241

Hamann W.-R., Gruschinske J., Kudritzki R.P., Simon K.P. 1981, A\&A 
104, 249

Heber U., Dreizler S., Werner K. 1993, Acta Astron. 43, 337




Rauch T. 1993, A\&A 276, 171

Zinn R.J., Newell E.B., Gibson J.B. 1971, A\&A 18, 390

\end{list}                                                             

\end{document}